# Cross-speaker Style Transfer with Prosody Bottleneck in Neural Speech Synthesis

*Shifeng Pan, Lei He*

Microsoft, China

`{peterpan, helei}@microsoft.com`

## Abstract

Cross-speaker style transfer is crucial to the applications of multi-style and expressive speech synthesis at scale. It does not require the target speakers to be experts in expressing all styles and to collect corresponding recordings for model training. However, the performances of existing style transfer methods are still far behind real application needs. The root causes are mainly twofold. Firstly, the style embedding extracted from single reference speech can hardly provide fine-grained and appropriate prosody information for arbitrary text to synthesize. Secondly, in these models the content/text, prosody, and speaker timbre are usually highly entangled, it's therefore not realistic to expect a satisfied result when freely combining these components, such as to transfer speaking style between speakers. In this paper, we propose a cross-speaker style transfer text-to-speech (TTS) model with explicit prosody bottleneck. The prosody bottleneck builds up the kernels accounting for speaking style robustly, and disentangles the prosody from content and speaker timbre, therefore guarantees high quality cross-speaker style transfer. Evaluation result shows the proposed method even achieves on-par performance with source speaker's speaker-dependent (SD) model in objective measurement of prosody, and significantly outperforms the cycle consistency and GMVAE-based baselines in objective and subjective evaluations.

## 1. Introduction

Recent advances in neural Text-to-speech (TTS) models [1-3] have greatly improved the quality of synthesized speech, which leads to a burst of TTS applications. There have been rising needs for multi-style (including emotions) TTS nowadays, to provide better listening experience for different content and scenarios. While the conventional way to expand the speaking style may still work, which is first collecting style recordings for the target speaker and then building a multi-style model, it's not scalable, especially for custom or personal TTS voices. Moreover, sometimes the target speaker may not perform well in certain styles, lowering the quality even in recording phase and resulting in unsatisfied synthesized speech. An alternative solution is transferring speaking style from cross-speaker source style recording to any target speaker, which is a more applicable way.

So far, towards transferring speaking style in TTS, the most related research topics are prosody transfer [4-7] and style transfer [8-12]. Prosody transfer studied in most papers relates to transferring fine-grained prosody from a text-paired reference speech to the synthesized speech. It's not applicable for a practical TTS system since the text paired reference cannot always be available. Style transfer studied in most papers relates to transferring "speaking style" from a reference speech to synthesized speech, where the reference can be text unpaired. The success of end-to-end (E2E) TTS model in generating highly natural speech encourages a similar belief that the style, which is difficult to clearly describe and model, is possible to be implicitly learned from the input mel-spectrogram and further transferred to synthesized speech, with certain style embedding network built upon an E2E TTS model. However, we observed that it's very difficult to force the network to sufficiently disentangle the style from content and timbre. What's more, many speaking styles are featured more by localized prosody variations or patterns instead of a global one, so a global embedding derived from a text-unpaired reference speech faces big challenges to recover fine-grained prosody on arbitrary target text while keeping sufficient stability. So far, the resulting transfer performance is far from good enough for application at scale.

In this work, firstly, we would like to define the basic specifications of cross-speaker style transfer TTS for real application as follows:

- Speaking style stands for how speech is expressed, such as in reading, chat, story-telling style, or in some emotional state such as happy and sad.
- No reference speech is needed, or at least text paired reference speech is not needed.
- Synthesized speech is basically in target speaker's timbre while with the cross-speaker source style.

Given the above requirements, we further propose a cross-speaker style transfer neural TTS, which is built upon a multi-speaker multi-style Transformer-based TTS model with explicit phone-level prosody bottleneck, consumed by the cross attention and decoder. The prosody bottleneck plays a key role in building up the kernels accounting for speaking style, disentangling the prosody from content and speaker timbre, and enabling a clear and easy solution for cross-speaker style transfer, as well as the controllability over prosody. The key idea can be formulated as : 1) prosody plays a key role in expression of speaking style and can be quantitatively described, 2) prosody can be robustly modelled given a certain amount of training data, 3) an encoder-attention-decoder based neural TTS, where the attention and decoder explicitly consume the input prosody, has good response to the input prosody, which in turn encourages the removal or disinterest of style related prosody information from text embedding and speaker embedding, and achieves good disentanglement between them, 4) by predicting the prosody of source speaker in source style and feeding it into the attention, which is further conditioned by the text and target speaker embedding, the cross-speaker source style can be robustly transferred to the target speaker. Evaluation result shows the proposed method achieves on-par performance with source speaker's SD model on objective measurement of prosody, and significantly better performance than baselines in style evaluation.[1] Here, the source speaker's SD model can be regarded as the upper boundary of style expression.

To the best of our knowledge, the contributions of this paper are as follows:
- For the first time, the basic specifications of cross-speaker style transfer TTS for real application are clearly defined.
- We reveal the intrinsic drawbacks of existing style transfer methods in building a cross-speaker style transfer TTS that is qualified for real application.
- We propose a new paradigm for cross-speaker speaker style transfer TTS, a multi-speaker multi-style transformer TTS with explicit prosody bottleneck, which significantly advances the performance of cross-speaker style transfer and allows flexible control over prosody as well.

## 2. Related Works

There have been some works in prosody transfer and style transfer fields in recently years, mostly built upon E2E TTS framework.

For prosody transfer, the goal is to transfer, or "copy", the prosody from the reference speech to the synthesized speech. The reference speech needs to have the same text as the target. In [4], the mel-spectrogram of the whole reference speech is encapsulated into one single embedding, where the whole prosody could be preserved, to some extent. The prosody can be well reproduced when target text is the same as the reference. In [5], a temporal structure is introduced into the embedding networks, which enables fine-grained prosody control of the synthesized speech. In [6], phone-level prosody features are extracted from reference and aggregated to the output of phone encoder, which enables a more accurate phone level prosody transfer and control. In [7], a frame-level reference encoder with a variational encoder and a temporal bottleneck encoder is adopted, which enables frame-level and robust prosody transfer, especially for cross-speaker prosody transfer. A key difference between the above works and the proposed method is that the prosody to transfer is not extracted from any reference speech but generated by the prosody bottleneck sub-network.

For style Transfer, in [8] "global style tokens" (GST) is proposed, which is a bank of embeddings jointly trained with Tacotron and works as intermediate anchors for transferring speaking style. In [9], variational autoencoder is introduced into E2E TTS, which can infer a more general and effective style embedding for style transfer. In [10], an unpaired path is introduced, where the reference speech has different text than the target, to address the mismatch between paired training and unpaired transfer in inference. All the above 3 works only focus on intra-speaker style transfer. While for cross-speaker style transfer, it's more challenging for that the mismatch from cross-speaker unparallel reference is more severe than an intra-speaker one. In [11], a so-called "cycle consistency" loss is introduced to constrain the unpaired output (cross-speaker transfer output) in training, which is measured by the difference of sentence-level embedding in latent space between the cross-speaker reference and the transferred output. In [12], the authors use Principal Component Analysis on style embeddings and use the first 3 components to do cross-speaker style transfer. The proposed method differs from [8-10] in that it is specially designed to address the challenges of cross-speaker style transfer. What's more, the proposed method achieves the transfer of style by retaining the phone-level prosody of source speaker in source style while preserving target speaker's timbre, which is more robust and fine-grained than all the above methods.

## 3. Proposed Model

The overall structure of the prosed model is illustrated in Figure 1, which is built upon the Transformer TTS [3].

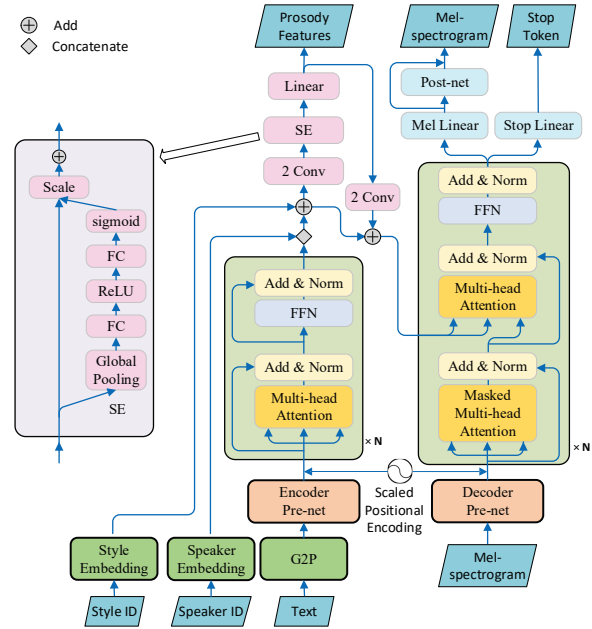

Figure 1: *Overall structure of the proposed model*

### 3.1. Transformer TTS

Compared with RNN-based model, transformer TTS has 2 major advantages. Firstly, the self-attention mechanism allows direct connection to global context, facilitating the modelling of long-range dependency. Secondly, it enables parallel training, which greatly reduces the training cost. The Transformer TTS takes text as input and output mel-spectrogram and stop token. The whole model mainly consists of a grapheme-to-phoneme (G2P) module, an encoder pre-net, a stack of encoder blocks, a decoder pre-net, a stack of decoder blocks, a post-net, and a stop token layer. The hyper parameters used in this work are the same as those in [3].

### 3.2. Multi-speaker multi-style Transformer TTS

The vanilla Transformer TTS, which is a single speaker mono style model, is further expanded with speaker and style embeddings to consume multi-speaker and multi-style data. Both the speaker and style embedding networks consist of a 128-dimensional look-up table. The speaker embedding is first broadcast-concatenated with the output of transformer encoder, the result of which is then projected back to 512-dimension. The style embedding is first projected to 512-dimension with tanh activation, then added to the previous speaker-combined encoder output. The result is referred to as speaker-style-combined encoder output hereafter.

---

[1]Audio samples available at
https://peterpanseu.github.io/index.html

### 3.3. Prosody bottleneck

E2E TTS omits some intermediate components of a classical TTS modelling flow. A powerful E2E network usually performs better than an ensemble of separately trained modules in overall naturalness, at the price of a highly entangled network and the difficulty in modification or manipulation. Cross-speaker style transfer on E2E TTS just suffers from such difficulty. To address this problem, we propose to introduce a prosody bottleneck sub-network into E2E TTS. As illustrated in Figure 1, the prosody bottleneck takes as input the speaker-style-combined encoder output and output phone-level prosody features, which can be viewed as prosody bottleneck of an E2E model. The prosody features used here include f0, voiced/unvoiced decision, duration, and energy. The prosody features are further added to the speaker-style-combined encoder output after going through a 2-layer CNN with 512 channels, forming a fully aggregated encoder output. The choosing of phone-level prosody is a tradeoff between the necessity of prosody disentangling and the loss to E2E benefit. As we will show later, it works quite well.

The prosody bottleneck network consists of a 2-layer CNN, followed by a squeeze-and-excitation (SE) block [13] and a linear layer. The SE block, as stated in [13], can adaptively recalibrate channel-wise feature responses by explicitly modelling interdependencies between channels, and produce significant performance improvements.

### 3.4. Training and generation

The training loss of the proposed model is shown in Eq. 1:

$$\mathcal{L} = \mathcal{L}_{spec} + \alpha \mathcal{L}_{stop} + \beta \mathcal{L}_{prosody} \quad (1)$$

, where $\mathcal{L}_{spec}$ is spectrum reconstruction loss, $\mathcal{L}_{stop}$ is stop token loss, $\mathcal{L}_{prosody}$ is prosody loss, and $\alpha$ and $\beta$ are the weights. $L^2$ loss is used for $\mathcal{L}_{spec}$ and $\mathcal{L}_{prosody}$, and cross-entropy is used for $\mathcal{L}_{stop}$.

In training, there's an option whether to feed the predicted prosody features to the cross-attention or the ground truth ones. We observe the former tends to yield a relatively richer prosody because of the fully joint training, while the latter yields a decoder which responds to the predicted prosody more faithfully.

Given source speaker $SpkID_{src}$, source style $StyID_{src}$, target speaker $SpkID_{tgt}$, and text, we do cross-speaker style transfer generation following below steps:

1) use $SpkID_{src}$, $StyID_{src}$, and text to generate prosody bottleneck, which corresponds to source speaker's phone-level prosody in source style given input text, then further go through the 2-layer CNN that follows;
2) use $SpkID_{tgt}$, $StyID_{src}$, and text to generate speaker-style-combined encoder output, simply reuse the text encoder output and style embedding generated in 1);
3) use the output of 1) and 2) to generate the fully aggregated encoder output, then go on with decoding.

## 4. Experiment

Two experiments are conducted in this paper. The first one (Exp-1) is to evaluate the performance of model trained from scratch with sufficient data, while the second one (Exp-2) is to study the performance of onboarding new source and target speakers with less data, given a pretrained source model.

In addition, the proposed model also allows flexible prosody control by directly modifying the phone-level prosody features. Due to the limited length of paper, we can only show this capability in the demo page. Readers are strongly encouraged to listen to the samples there.

### 4.1. Experimental setup

In both experiments, two internal female American speakers are involved, where the source speaker (Speaker A) is generally in chat style with multiple emotions, and the target speaker (Speaker B) is in reading style with neutral emotion. The distribution of training data is listed in Table 1. The data is recorded with 48kHz/24bit, and down sampled to 16kHz/16bit and 24kHz/16bit for acoustic model and vocoder training, respectively. 80-dimensional mel-spectrogram is extracted from the 16kHz waveform. An external aligner is used to get phone boundary, then phone-level prosody features are extracted and normalized globally. The vocoder used in this experiment is Microsoft HiFiNet [14], which achieves human-parity audio fidelity.

Table 1: *Training data distribution.*

| Experiment | Speaker | Neutral | Happy | Sad | Angry |
|---|---|---|---|---|---|
| 1 | A | 10k | 4k | 4k | 4k |
| 1 | B | 10k | - | - | - |
| 2 | A | 500 | - | - | - |
| 2 | B | 500 | - | - | - |

In Exp-1, the model is trained from scratch and the goal is to transfer the 3 emotions of Speaker A to Speaker B. In Exp-2, firstly, a multi-speaker multi-style source model is trained, which involves 80 American speakers, 10 styles, and 150 hours in total, with Speaker A and B excluded. Then, model refinement is performed with Speaker A and B's data on top of the above well-trained source model. During refining, we tried different strategies, such as refining full model, whole encoder (including speaker, style, prosody embeddings), encoder plus cross attention, etc. The result reported here for Exp-2 is based on the model with whole encoder refined.

The models evaluated in this work are listed below, all of which are based on Transformer TTS framework:

- **Spk-A_SD**: Speaker A's SD model, viewed as the upper boundary for style evaluation.
- **Spk-B_SD**: Speaker B's SD model, viewed as the lower boundary for style evaluation.
- **Spk-B_Trans_CC**: A Transformer TTS version of the cycle consistency loss enhanced method in [11].
- **Spk-B_Trans_GMVAE**: Gaussian Mixture Variational Autoencoder (GMVAE) based style transfer model. Similar with [9], variational inference is introduced into Transformer TTS, but Gaussian mixture model is used instead of single Gaussian for prior distribution. The mixtures are explicitly tied to each style. In inference, the mean of the mixture corresponding to target style is used.
- **Spk-B_Trans_Pros**: The proposed model.

### 4.2. Test metrics

The test metrics used in the evaluation are listed below:

- **Prosody Measurement**: Phone-level prosody correlation to source style recording, include LogF0 (Lf0), Duration (Dur) and Energy. Lf0 RMSE is also measured here.

- **Speaker Classification**: To verify if the transferred voice preservers target speaker's timbre. A simple classifier with 6-layer CNN + GRU (similar with the reference encoder in [4]) and a FC layer is trained to do the test.
- **Mean Opinion Score (MOS)**: To verify the naturalness.
- **AB Preference (ABX)**: To verify which one's style is more similar with the source style recording.
- **Emotion Perception**: A subjective emotion perception test. The judger is asked to select one from 5 options, including the 3 emotions, neural, and none, according to his/her perception on the test case.

A small test set of Speaker A is excluded from training set, which provides test script for test case generation and ground truth recording for each of the tests. All the subjective tests are conducted by crowd-sourced native judgers. Usually, each judger is limited with 50 judgements in one test, to avoid listening fatigue. Test settings are listed in Table 2.

Table 2: *Test settings*.

| Exp | Metrics | #Test case | #Judges/case |
|---|---|---|---|
| 1 & 2 | Prosody correlation | 50 | - |
| 1 & 2 | Speaker Classification | 50 | - |
| 1 | MOS | 20/emotion | 20 |
| 1 | ABX | 30/emotion | 9 |
| 1 | Emotion Perception | 30/emotion | 9 |
| 2 | MOS | 30 | 20 |
| 2 | ABX | 50 | 9 |

### 4.3. Results

As shown in Table 3, the general naturalness of all models is good, with MOS > 4.0. There's no statistical difference among the models except for **Spk-B_SD** in Exp-1, which might be caused by the lack of emotional tone.

The prosody measurements in Table 4 and 5 show that the proposed method is significantly better than all other transfer models and close to the upper bar **Spk-A_SD**. The lf0 measurements of **Spk-B_Trans_Pros** are even slightly better than **Spk-A_SD**. We observe the Lf0 dynamics of **Spk-A_SD** is bigger than **Spk-B_Trans_Pros**, which tends to yield bigger impact when there's some errors. Table 6 shows that the speaker classification results in Exp-1 for the 3 transferred voices are a little bit lower, around 91%-93%, while the result in Exp-2 is very good. By listening to the failed cases, we don't see the timbre get obviously changed. We think it might be caused by the classification model itself, which didn't see any emotional samples of Spk-B in training, but Spk-A's instead. Given that the emotional speech is quite different from the neural speech, the generalization ability of the classification model is likely to be limited for the emotion transferred voices. We'll verify this hypothesis in future.

The subjective tests for style evaluation are shown in Table 7 and 8. In Exp-1, the proposed model is significantly preferred than cycle consistency and VAE based transfer methods, with **53.9%** and **34.3%** absolute gains, respectively.

Table 3: *MOS*.

| Model | Exp-1 | Exp-2 |
|---|---|---|
| Recording_Spk-B | 4.29 ± 0.05 | 4.29 ± 0.07 |
| Spk-B_SD | 4.01 ± 0.05 | 4.06 ± 0.07 |
| Spk-B_Trans_CC | **4.13** ± 0.04 | - |
| Spk-B_Trans_GMVAE | 4.08 ± 0.04 | 4.08 ± 0.06 |
| Spk-B_Trans_Pros | 4.07 ± 0.05 | **4.11** ± 0.07 |

Table 4: *Prosody Measurement of Exp-1*.

| Model | Lf0_Corr | Dur_Corr | Energy_Corr | Lf0_RMSE |
|---|---|---|---|---|
| Spk-A_SD | 0.425 | **0.848** | **0.915** | 0.255 |
| Spk-B_SD | 0.226 | 0.749 | 0.659 | 0.302 |
| Spk-B_Trans_CC | 0.230 | 0.794 | 0.884 | 0.284 |
| Spk-B_Trans_GMVAE | 0.382 | 0.837 | 0.907 | 0.262 |
| Spk-B_Trans_Pros | **0.439** | 0.844 | 0.893 | **0.237** |

Table 5: *Prosody Measurement of Exp-2*.

| Model | Lf0_Corr | Dur_Corr | Energy_Corr | Lf0_RMSE |
|---|---|---|---|---|
| Spk-A_SD | 0.638 | **0.843** | **0.931** | 0.175 |
| Spk-B_SD | 0.402 | 0.786 | 0.892 | 0.222 |
| Spk-B_Trans_GMVAE | 0.503 | 0.821 | 0.91 | 0.198 |
| Spk-B_Trans_Pros | **0.66** | 0.842 | 0.917 | **0.156** |

Table 6: *Speaker Classification Accuracy (%)*.

| Model | Exp-1 | Exp-2 |
|---|---|---|
| Spk-A_SD | 100 | 100 |
| Spk-B_SD | 97.3 | 100 |
| Spk-B_Trans_CC | 92.7 | - |
| Spk-B_Trans_GMVAE | 91.3 | 100 |
| Spk-B_Trans_Pros | 91.3 | 98 |

Similar ranking can be found in emotion perception test, where the proposed model has **11.7%** and **8.3%** absolution gains, respectively. In Exp-2, the proposed method is also significantly preferred than VAE based method, with **33.6%** absolute gain.

Table 7: *ABX (%)*.

| Exp | Spk-B_SD | Spk-B_Trans_CC | Spk-B_Trans_GMVAE | Spk-B_Trans_Pros | Equal |
|---|---|---|---|---|---|
| 1 | 18.4 | | | **77.7** | 4.0 |
| 1 | | 27.3 | | **62.2** | 10.6 |
| 1 | | | 28.9 | **63.2** | 7.9 |
| 2 | 29.3 | | | **66.3** | 4.3 |
| 2 | | | 30.3 | **63.9** | 5.8 |

Table 8: Emotion *Perception Accuracy of Exp-1 (%)*.

| Model | Happy | Angry | Sad | Average |
|---|---|---|---|---|
| Spk-A_Recording | 60.74 | 67.78 | 75.19 | 67.9 |
| Spk-A_SD | **58.52** | 60.37 | 82.59 | 67.16 |
| Spk-B_SD | 26.30 | 33.70 | 20.00 | 26.67 |
| Spk-B_Trans_CC | 57.41 | 37.78 | 57.04 | 50.74 |
| Spk-B_Trans_GMVAE | 51.48 | 31.85 | 54.07 | 45.80 |
| Spk-B_Trans_Pros | **58.52** | 41.85 | 71.11 | 57.16 |

## 5. Conclusions

In this paper, we analyze the drawbacks of current prosody transfer and style transfer methods in building a cross-speaker style transfer TTS that is qualified for real application. A multi-speaker multi-style transformer TTS with explicit prosody bottleneck is then proposed. Evaluation result shows it significantly advances the performance of cross-speaker style transfer. In addition, it has good capability of prosody control.